\documentclass[prb,aps,twocolumn,showpacs]{revtex4}
\usepackage{dcolumn}
\usepackage{graphicx}
\usepackage{amssymb}
\begin{document}

\title{Magnetoresistive junctions based on epitaxial graphene and hexagonal boron nitride}

\author{Oleg V. Yazyev}
\altaffiliation{Present address: Department of Physics, University of 
California, Berkeley, CA 94720, USA; E-mail: yazyev@civet.berkeley.edu}
\affiliation{Institute of Theoretical Physics, Ecole Polytechnique 
F\'ed\'erale de Lausanne (EPFL), CH-1015 Lausanne, Switzerland}
\affiliation{Institut Romand de Recherche Num\'erique en Physique
des Mat\'eriaux (IRRMA), CH-1015 Lausanne, Switzerland}
\author{Alfredo Pasquarello}
\affiliation{Institute of Theoretical Physics, Ecole Polytechnique 
F\'ed\'erale de Lausanne (EPFL), CH-1015 Lausanne, Switzerland}
\affiliation{Institut Romand de Recherche Num\'erique en Physique
des Mat\'eriaux (IRRMA), CH-1015 Lausanne, Switzerland}

\date{\today}

\begin{abstract}
We propose monolayer epitaxial graphene and hexagonal boron nitride 
($h$-BN) as ultimate thickness covalent spacers for 
magnetoresistive junctions. Using a first-principles approach, we 
investigate the structural, magnetic and spin transport properties 
of such junctions based on structurally well defined interfaces with 
(111) fcc or (0001) hcp ferromagnetic transition metals. We find  
low resistance area products, strong exchange couplings across the 
interface, and magnetoresistance ratios exceeding 100\% for certain 
chemical compositions. These properties can be fine tuned, making 
the proposed junctions attractive for nanoscale spintronics applications.
\end{abstract}

                
\maketitle

\section{INTRODUCTION}

Graphene, a recently discovered two-dimensional form of carbon, has attracted
unrivaled attention due to its unique physical properties and potential 
applications in electronics.\cite{Katsnelson07,Geim07} This nanomaterial is 
particularly promising for the field of spintronics, which exploits both the 
spin and the charge of electrons.\cite{Son06,Tombros07,Yazyev08,Munoz-Rojas09,Yazyev08c,Yazyev09,Yazyev08b} 
One fundamental spintronic effect is the magnetoresistance, the change in electric
resistance as a function of the relative orientation, either parallel or 
antiparallel, of the magnetization of two ferromagnetic layers separated by 
a nonmagnetic spacer layer.\cite{Heiliger06} Achieving {\it high} magnetoresistance 
ratios while keeping reasonably {\it low} electric resistance is crucial for many 
technological applications.\cite{Chappert07} However, reaching this goal is 
currently hindered by material-specific restrictions such as the inability 
of producing well-ordered ferromagnet/spacer interfaces.\cite{Yuasa04,Heiliger06b} 

Semimetallic graphene and its insulating counterpart, isostructural hexagonal 
boron nitride ($h$-BN), are promising spacers as epitaxial monolayers of these 
materials can be grown by means of chemical vapor deposition (CVD) on a broad 
variety of metallic substrates.\cite{Oshima97,Berner07,Coraux08,deParga08,Sutter08,Martoccia08}
The quality of such epitaxial monolayers is very high and the covalent bonding 
network of both graphene and $h$-BN is perfectly preserved upon the bonding to 
the substrate. Moreover, the growth of graphene and $h$-BN on fcc(111) 
and hcp(0001) surfaces of ferromagnetic Co and Ni results in commensurate 
epitaxial layers due to the closely matching lattice constants.\cite{Oshima97} 
This has led to a theoretical prediction of perfect spin filtering and, thus,
to extremely high magnetoresistance ratios in such junctions based on 
multilayer graphene ($\ge$4 layers) and graphite.\cite{Karpan07} 
However, the CVD growth on crystalline surfaces is self-inhibiting, that is only 
one epitaxial layer can be grown. The deposition of ferromagnetic nanoparticles 
on top of epitaxial $h$-BN has also been demonstrated.\cite{Auwarter02,Zhang08} 
These interfaces further offer the opportunity of fine tuning their properties 
through the intercalation of other metals, such as Fe,\cite{Dedkov08} 
Cu\cite{Dedkov01} and Au.\cite{Varykhalov08} 

In this work, we suggest the use of monolayer graphene and $h$-BN as {\it covalently} bonded
spacer layers of minimal thickness in magnetoresistive junctions. 
Through first principles calculations we study the structural, magnetic and spin transport 
properties of such junctions based on first-row ferromagnetic transition metals:
natural hcp and fcc Co, fcc Ni, as well as intercalated fcc Fe. We show that 
the proposed magnetoresistive junctions realize low electric resistances, 
strong interlayer exchange couplings, and magnetoresistance ratios exceeding 
100\% for certain chemical compositions.

This paper is organized as follows. In Sec.\ \ref{sec2} we describe our computational
methodology, including the first-principles approach to electronic transport. 
In Sec.\ \ref{sec3} we report the atomic structure and electronic properties
of the considered magnetoresistive junctions. Particular attention is devoted
to the interlayer exchange couplings. The results of electronic 
transport calculations are discussed in Sec.\ \ref{sec4}. Section\ \ref{sec5}
concludes our work.

%
%

\section{COMPUTATIONAL METHODS}\label{sec2}

The electronic and atomic structure calculations were performed using the 
\textsc{pwscf} plane-wave pseudopotential code of the \textsc{quantum-espresso}
distribution.\cite{QE} To achieve a good description of atomic structures, 
interlayer exchange couplings and spin transport properties, we chose the 
Perdew-Burke-Ernzerhof exchange-correlation density functional.\cite{Perdew96} 
Ultrasoft pseudopotentials were used to describe core-valence interactions.\cite{Vanderbilt90} 
The valence wave functions and the electron density were described by plane-wave 
basis sets with kinetic energy cutoffs of 25~Ry and 250~Ry, respectively.\cite{Pasquarello92}
The atomic structure of the magnetoresistive junctions considered in 
our work is illustrated in Fig.\ \ref{fig1}(a).
Our investigation is restricted to only symmetric junctions, 
i.e.\ with the same metal on both sides of the spacer layer. Each ferromagnetic 
layer consisted of six atomic planes. The solutions for parallel and antiparallel 
relative spin orientations  of these two layers were obtained by specifying
appropriate initial orientations of the magnetic moments. The lateral unit cell 
of the studied interfaces is shown in Fig.~\ref{fig1}(b). We considered bound
configurations and determined the lowest-energy structures through the relaxation
of atomic positions. For these configurations, we performed quantum transport 
calculations in the current perpendicular to plane configuration using 
the \textsc{pwcond} code\cite{Smogunov04} 
of the same package. The scattering region included the spacer monolayer and three 
adjacent monolayers of metal on both sides. We use the optimistic definition of 
the magnetoresistance ratio:
\begin{equation}
 {\rm MR} = \frac{G_{\uparrow\uparrow}+G_{\downarrow\downarrow}-2G_{\uparrow\downarrow}}{2G_{\uparrow\downarrow}}\times 100\%.
\end{equation}
The spin-resolved quantum conductances $G_\sigma$
for parallel ($\sigma = \uparrow\uparrow, \downarrow\downarrow$ for majority and minority spins, 
respectively) and antiparallel ($\sigma = \uparrow\downarrow$) configurations were 
calculated by integrating the corresponding ${\mathbf k}_{||}$-dependent transmission 
probabilities $T^\sigma_{{\mathbf k}||}$ evaluated on a uniform grid of 64$\times$64 
${\mathbf k}$-points in the two-dimensional Brillouin zone.

\section{ATOMIC AND ELECTRONIC STRUCTURE}\label{sec3}

%
%

\begin{figure}
\includegraphics[width=8.5cm]{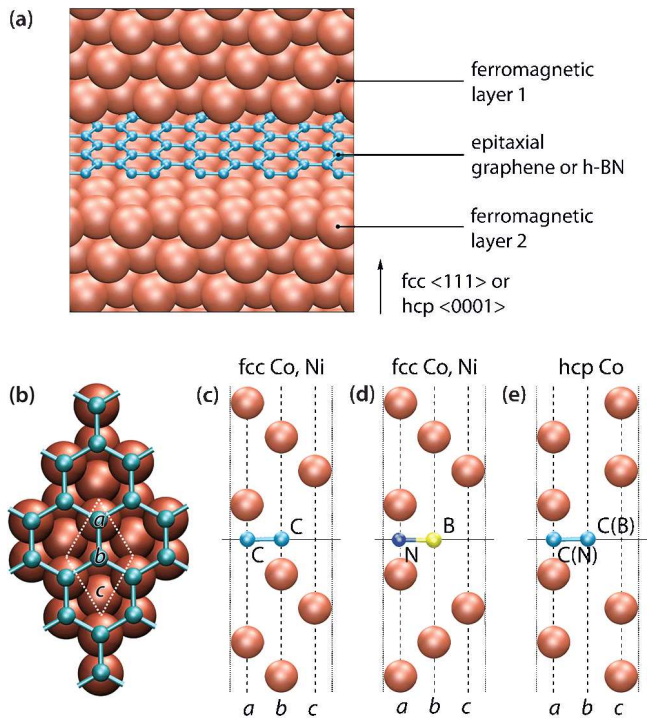}
\caption{\label{fig1} (Color online) 
(a) Representation of the atomic structure of magnetoresistive junctions
based on epitaxial monolayer graphene and $h$-BN.
(b) Top-view of graphene on (111) surface of fcc Co or Ni. The two-dimensional 
unit cell is indicated by dotted lines and the principal atomic positions are 
labeled. (c)--(e) Side-views along the longest unit cell diagonal of the 
lowest-energy interfaces formed by Co (hcp and fcc) and Ni (fcc) in combination 
with either monolayer graphene or $h$-BN.
}
\end{figure}

To determine the lowest energy structures of the junctions, we carried out 
structural relaxations for all possible stacking orders of the atomic planes 
in the vicinity of the spacer layer. The corresponding structures are shown in 
Fig.~\ref{fig1} for Co and Ni based junctions and summarized in Table~\ref{tab1}
for all investigated chemical compositions. We find that both graphene (GR) 
and $h$-BN bound to the transition metals (TM) display short metal-carbon and 
metal-nitrogen distances (2.19--2.45~\AA) comparable to the sum of the corresponding 
covalent bond radii. The thickness of the spacer layer is thus comparable to that
of a single atomic plane of the ferromagnetic metal. For the Ni$|$GR$|$Ni(fcc)
junction we find a Ni--C distance of 2.19~\AA, which is close to 2.18~\AA\ 
calculated for the graphene chemisorbed on the Ni(111) surface (i.e.\ Ni$|$GR
system). The latter value is in good agreement with the experimental value of 
2.16$\pm$0.07~\AA.\cite{Oshima97} 

%
%

\begin{figure}[b]
\includegraphics[width=8.5cm]{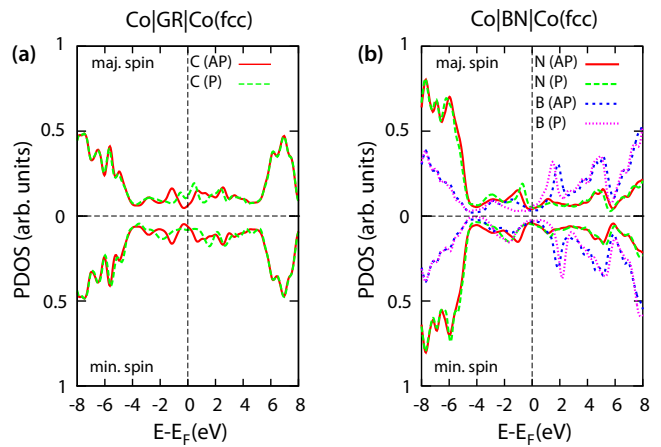}
\caption{\label{fig2} (Color online) 
Spin-resolved projected density of states (PDOS) onto the atoms of the spacer 
layer, either (a) graphene or (b) $h$-BN, in the fcc Co junctions in their 
parallel (P) and antiparallel (AP) configurations. The majority and minority
spin labels refer to the parallel configuration; in the antiparallel arrangement
the spin channels are equivalent. 
}
\end{figure}

We now turn to the electronic structure of these magnetoresistive junctions. 
Figure~\ref{fig2} shows the spin-resolved projected density of states (PDOS) 
onto the light atoms (B, C and N) of the fcc Co junctions. 
One can see that the characteristic ``Dirac cone'' density of states of 
the free-standing graphene is not preserved upon the formation of the 
Co$|$GR$|$Co(fcc) interface [Fig.~\ref{fig2}(a)]. This is consistent with 
the theoretically predicted\cite{Karpan07} and experimentally 
observed\cite{Gruneis08} strong hybridization between the electronic 
states of graphene and of the TM surface. Similarly, both B and N centered 
states fill the band gap of the insulating $h$-BN [Fig.~\ref{fig2}(b)]. 
In both cases, we very similarly find significant contributions of the epitaxial layer 
states to the density of states at the Fermi level. In the parallel
(antiparallel) configuration of the graphene based junction, the induced
magnetic moments on the carbon atoms in the unit cell are $-$0.005$\mu_{\rm B}$
(0.081$\mu_{\rm B}$ and $-$0.081$\mu_{\rm B}$). In the parallel configuration
of the $h$-BN junction, the induced magnetic moments of N and B atoms are 
0.029$\mu_{\rm B}$ and $-$0.065$\mu_{\rm B}$, respectively. In the 
antiparallel configuration both vanish by symmetry. 

\begin{table*}
\caption{\label{tab1} 
Lowest energy structures, interlayer exchange couplings ($\Delta E = E_{\rm P}-E_{\rm AP}$), 
spin-resolved quantum conductances ($G_{\uparrow\uparrow}$, $G_{\downarrow\downarrow}$
and $G_{\uparrow\downarrow}$) and magnetoresistance ratios (MR) for the 
discussed graphene (GR) and $h$-BN junctions. The notation for the stacking 
order is identical to the one in Fig.\ \ref{fig1}. The values of quantum 
conductances and interlayer exchange couplings are given per unit cell area.
}
\begin{ruledtabular}
\begin{tabular}{lcccccc}
~~~~junction & stacking &  $\Delta E$ & $G_{\uparrow\uparrow}$ & $G_{\downarrow\downarrow}$ & $G_{\uparrow\downarrow}$ & MR   \\
             &  order   &   (meV)     & ($e^2/h$)              & ($e^2/h$)                  & ($e^2/h$)                & (\%) \\ 
\hline
Fe$|$GR$|$Fe(fcc) & $cba$$|$$bac$ & 79 & 0.334 & 0.440 & 0.240 & 61  \\
Fe$|$BN$|$Fe(fcc) & $cba$$|$$abc$ & 63 & 0.256 & 0.297 & 0.111 & 149 \\
Co$|$GR$|$Co(fcc) & $bca$$|$$bca$ & 91 & 0.317 & 0.427 & 0.232 & 60  \\
Co$|$BN$|$Co(fcc) & $bca$$|$$acb$ & 46 & 0.263 & 0.268 & 0.210 & 26  \\
Ni$|$GR$|$Ni(fcc) & $bca$$|$$bca$ & $-$18 & 0.352 & 0.587 & 0.402 & 17  \\
Ni$|$BN$|$Ni(fcc) & $bca$$|$$acb$ & $-$3  & 0.207 & 0.722 & 0.299 & 55  \\
Co$|$GR$|$Co(hcp) & $aca$$|$$aca$ & 29 & 0.241 & 0.278 & 0.140 & 86  \\
Co$|$BN$|$Co(hcp) & $aca$$|$$aca$ & 44 & 0.222 & 0.241 & 0.140 & 66  \\
\end{tabular}
\end{ruledtabular}
\end{table*}

The interlayer exchange coupling, the difference $\Delta E = E_{\rm P}-E_{\rm AP}$ 
between the energies of parallel and antiparallel configurations, is a manifestation 
of the superexchange mechanism. It achieves rather high values [cf.\ Table\ \ref{tab1}]
due to the ultimate thickness of the spacer layer. For Fe and Co, the antiparallel 
configuration is energetically favored. 
On the contrary, the parallel configuration is preferred for Ni. This intriguing 
crossover provides opportunity for fine tuning the interlayer exchange by varying 
the chemical composition of the ferromagnetic layers.

%
%

\section{ELECTRONIC TRANSPORT}\label{sec4}

\subsection{Role of spacer material}

\begin{figure}
\includegraphics[width=8.5cm]{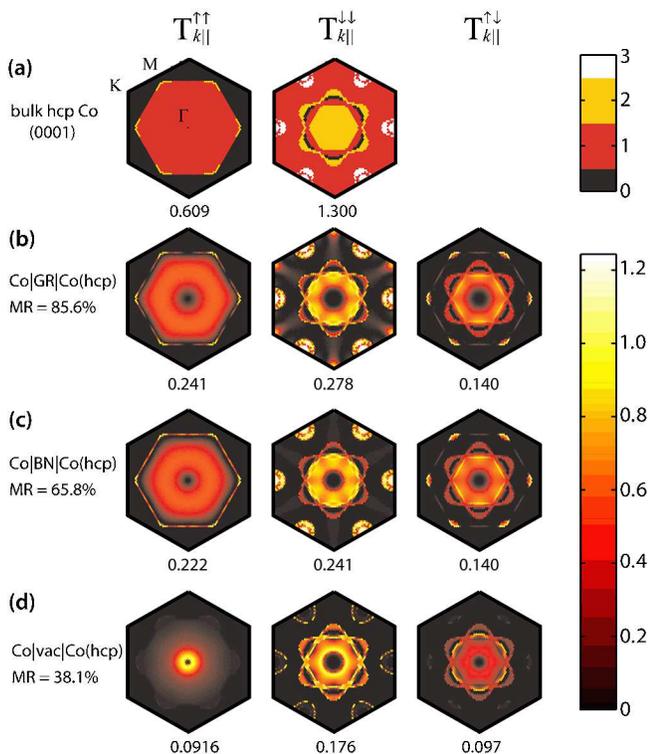}
\caption{\label{fig3} (Color online) 
${\mathbf k}_{||}$-Resolved conductance per unit cell (in units of $e^2/h$)
through (a) bulk hcp Co along the (0001) direction, (b) Co$|$GR$|$Co(hcp) 
and (c) Co$|$BN$|$Co(hcp) junctions, and (d) a vacuum layer of equivalent 
thickness. The columns correspond to majority and minority spin channels of 
the parallel configuration, and to one of the equivalent spin channels of 
the antiparallel configuration, respectively. Labels indicate the 
total conductances per unit cell area.
}
\end{figure}

To understand the calculated quantum conductances and the resulting 
magnetoresistance ratios [cf.\ Table~\ref{tab1}], we analyzed the 
${\mathbf k}_{||}$-resolved transmission probabilities. First, we studied 
the effect of the spacer layer in hcp Co junctions which have the same 
lowest energy structure for both graphene and $h$-BN [Fig.~\ref{fig3}].
We found that both systems show strikingly similar ${\mathbf k}_{||}$-resolved 
transmission probability maps [compare Figs.~\ref{fig3}(b) and \ref{fig3}(c)]
and consequently quantum conductances. The $T^{\uparrow\uparrow}_{{\mathbf k}||}$ 
and $T^{\downarrow\downarrow}_{{\mathbf k}||}$ maps reveal major features of 
the projected hcp Co Fermi surfaces for the free-electron-like majority spin 
and mostly $d$-symmetry minority spin electrons [Fig.~\ref{fig3}(a)], which are
relevant to the quantum conductances of bulk metals.\cite{Schep98,Zwierzycki08} 
The total transmission probabilities of the junctions in the parallel 
configurations constitute $\sim$40\% and $\sim$20\% of the Sharvin conductances\cite{Sharvin65} 
of bulk hcp Co along the (0001) direction. The quantum 
conductances in the antiparallel configuration are mostly determined by the 
overlap of $T^{\uparrow\uparrow}_{{\mathbf k}||}$ and $T^{\downarrow\downarrow}_{{\mathbf k}||}$.
Their values are consequently lower ($G_{\uparrow\downarrow}=0.140$~$e^2/h$ per unit 
cell for both spacer materials). The resulting magnetoresistance ratios are
86\% and 66\% for graphene and $h$-BN junctions. Thus, in the regime of 
ultimate thickness the transport properties are largely independent of the 
electronic structure differences of the two spacer materials. The role of a single 
layer of covalent spacer material consists in fixing a certain stacking order 
at the interface and in providing a medium for the abrupt change of magnetization 
in the antiparallel configuration. Due to the
metallic nature of the spacer layers [cf.\ Fig.\ \ref{fig2}] such junctions
possess low resistance area products ($<$3$\times$10$^{-15}$~$\Omega$m$^2$) 
which makes them suitable for nanoscale spintronics applications such as 
the magnetic random access memories and spin transfer nano-oscillators.
We classify the present systems as {\it giant magnetoresistance} (GMR) junctions. 
This contrasts to the spin transport through a vacuum gap of the same thickness 
which shows $T^\sigma_{{\mathbf k}||}$ decaying with $|{\mathbf k}_{||}|$ 
[Fig.~\ref{fig3}(d)], a characteristic feature of tunneling.\cite{Belashchenko04} 
The magnetoresistance ratio is about twice smaller (38\%) in the case of tunneling 
through a vacuum gap. This allows us to conclude that in the limit of ultimate 
thickness the GMR effect is more efficient than the tunneling magnetoresistance. 
 
\subsection{Role of ferromagnetic layers}

\begin{figure}[b]
\includegraphics[width=8.5cm]{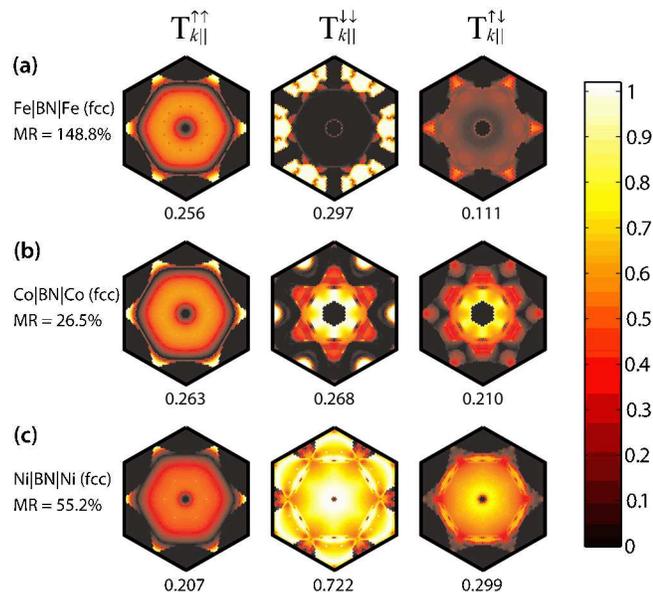}
\caption{\label{fig4} (Color online) 
${\mathbf k}_{||}$-Resolved conductance per unit cell (in units of $e^2/h$)
through fcc (a) Fe, (b) Co and (c) Ni junctions based on monolayer $h$-BN. 
The columns correspond to majority and minority spin channels of the parallel 
configuration, and to one of the equivalent spin channels of the antiparallel 
configuration, respectively. Labels indicate the corresponding total 
conductances per unit cell area.
}
\end{figure}

Next, we studied the dependence of transport properties on the ferromagnetic 
metal by considering fcc Fe, Co and Ni junctions in combination with $h$-BN.
For all three metals, the majority spin transmission in the parallel
configuration, $T^{\uparrow\uparrow}_{{\mathbf k}||}$, undergoes little 
change along the Fe-Co-Ni series [Fig.~\ref{fig4}]. This behavior stems from 
the similarity of the corresponding majority spin Fermi surfaces of the bulk 
metals, which are formed by partially filled $s$ bands. However, much larger 
differences are found for $T^{\downarrow\downarrow}_{{\mathbf k}||}$ involving
the minority electrons. These reflect the drastically different Fermi surfaces
resulting from the interplay between $s$ and $d$ states. The increase of 
$G_{\downarrow\downarrow}$ along the series can be attributed to the decrease 
of hybridization between $s$ and $d$ electrons upon the increase of $d$ band 
filling:\cite{Mazin99}
in general, the free-electron-like $s$ states show higher
transmission probabilities. The $G_{\uparrow\downarrow}$ values are again
determined by the overlap of $T^{\uparrow\uparrow}_{{\mathbf k}||}$ and 
$T^{\downarrow\downarrow}_{{\mathbf k}||}$ and tend to increase along the 
series. For the Fe$|$BN$|$Fe(fcc) junction we find a magnetoresistance ratio 
of 150\%, the largest value among the compositions studied. 

Further search of magnetoresistive junctions with improved characteristics
may consist in exploring asymmetric junctions and the intercalation of some 
other chemical elements at the interfaces. We here demonstrate the second
possibility. It has been suggested that the incorporation of submonolayer 
quantities of Cu at the TM$|$GR$_n$ interface would reduce undesired 
hybridization between the states of graphene and of the metal surface at 
the price of substantially decreasing the magnetoresistance ratios.\cite{Karpan07} 
However, we find that the decoupling of the spacer layer from the metal surface 
does not necessarily imply the loss of magnetoresistance. This can be achieved 
by intercalating the metals from the middle of the transition metals series, 
e.g.\ Mn, which show reduced binding to carbon $\pi$ systems.\cite{Pandey01} 
Indeed, in the intercalated CoMn(1\ ML)$|$GR$|$Mn(1\ ML)Co(hcp) junction the Mn--C 
distance increases to 2.95~\AA\ and the interlayer exchange coupling decreases
to 10~meV (to be compared with 29~meV for Co$|$GR$|$Co(hcp), cf.\ Table~\ref{tab1}).
Concurrently, the magnetoresistance ratio raises from 86\% to 127\%. The Mn 
layer is strongly spin polarized and antiferromagnetically coupled to hcp Co. 

%
%

\section{CONCLUSIONS}\label{sec5}

In conclusion, we propose epitaxially grown monolayer graphene and $h$-BN as 
ultimate thickness covalent spacers in transition metal based magnetoresistive 
junctions. Such junctions display well-ordered interfaces and can be produced 
through existing manufacturing processes. Their physical properties can be 
fine tuned in a broad range by varying the chemical composition. These systems 
show low resistance area products and typical GMR behavior with magnetoresistance 
ratios exceeding 100\% for certain compositions. Both ferromagnetic and 
antiferromagnetic interlayer exchange couplings are found. These properties make 
the proposed junctions attractive for spintronics applications such as the 
magnetic random access memories and spin transfer nano-oscillators.

%
%

\section*{ACKNOWLEDGMENT}

We acknowledge fruitful discussions with H.~Brune, P.~J.~Kelly and S.~Rusponi.  
We would like to thank A.~Smogunov for his help with the \textsc{pwcond} code. 
The calculations were performed at the CSCS.

\end{document}